\pdfoutput=1
\documentclass[12pt,a4paper]{article}

\usepackage{ifthen} 
\newboolean{pdflatex}
\setboolean{pdflatex}{true} 

\newboolean{articletitles}
\setboolean{articletitles}{true} 

\newboolean{uprightparticles}
\setboolean{uprightparticles}{false} 

\newboolean{inbibliography}
\setboolean{inbibliography}{false} 

\def\paperauthors{LHCb collaboration} 
\def\papercopyright{\the\year\ CERN for the benefit of the LHCb collaboration} 
\def\paperlicence{CC BY 4.0 licence}
\def\paperlicenceurl{https://creativecommons.org/licenses/by/4.0/}

\usepackage[top=1in, bottom=1.25in, left=1in, right=1in]{geometry}

\usepackage{microtype}
\usepackage{lineno}  
\usepackage{xspace} 
\usepackage{caption} 

\usepackage{graphicx}  
\usepackage{color}
\usepackage{colortbl}
\graphicspath{{./figs/}} 

\usepackage{amsmath} 
\usepackage{amssymb}
\usepackage{amsfonts}
\usepackage{upgreek} 

\newcommand*\patchAmsMathEnvironmentForLineno[1]{%
\expandafter\let\csname old#1\expandafter\endcsname\csname #1\endcsname
\expandafter\let\csname oldend#1\expandafter\endcsname\csname
end#1\endcsname
 \renewenvironment{#1}%
   {\linenomath\csname old#1\endcsname}%
   {\csname oldend#1\endcsname\endlinenomath}%
}
\newcommand*\patchBothAmsMathEnvironmentsForLineno[1]{%
  \patchAmsMathEnvironmentForLineno{#1}%
  \patchAmsMathEnvironmentForLineno{#1*}%
}
\AtBeginDocument{%
\patchBothAmsMathEnvironmentsForLineno{equation}%
\patchBothAmsMathEnvironmentsForLineno{align}%
\patchBothAmsMathEnvironmentsForLineno{flalign}%
\patchBothAmsMathEnvironmentsForLineno{alignat}%
\patchBothAmsMathEnvironmentsForLineno{gather}%
\patchBothAmsMathEnvironmentsForLineno{multline}%
\patchBothAmsMathEnvironmentsForLineno{eqnarray}%
}

\usepackage{hyperref}    
\usepackage[all]{hypcap} 

\usepackage{xspace} 
\usepackage{upgreek}

\def\lhcb   {\mbox{LHCb}\xspace}

\def\MagUp {\mbox{\em Mag\kern -0.05em Up}\xspace}

\ifthenelse{\boolean{uprightparticles}}%
{
 
 \def\Pgamma      {\ensuremath{\upgamma}\xspace}

 \def\Pmu         {\ensuremath{\upmu}\xspace}                 
 \def\Pnu         {\ensuremath{\upnu}\xspace}                 
                  
 \def\Ppi         {\ensuremath{\uppi}\xspace}

 \def\Pphi        {\ensuremath{\upphi}\xspace}

 \def\Ppsi        {\ensuremath{\uppsi}\xspace}

 \def\PDelta      {\ensuremath{\Delta}\xspace}                 
 \def\PXi         {\ensuremath{\Xi}\xspace}                 
 \def\PLambda     {\ensuremath{\Lambda}\xspace}                 
 \def\PSigma      {\ensuremath{\Sigma}\xspace}                 
 \def\POmega      {\ensuremath{\Omega}\xspace}                 
 \def\PUpsilon    {\ensuremath{\Upsilon}\xspace}

 \def\PB      {\ensuremath{\mathrm{B}}\xspace}                 
                  
 \def\PD      {\ensuremath{\mathrm{D}}\xspace}

 \def\PJ      {\ensuremath{\mathrm{J}}\xspace}                 
 \def\PK      {\ensuremath{\mathrm{K}}\xspace}

 \def\Pb      {\ensuremath{\mathrm{b}}\xspace}                 
 \def\Pc      {\ensuremath{\mathrm{c}}\xspace}

 \def\Ph      {\ensuremath{\mathrm{h}}\xspace}                 
 \def\Pi      {\ensuremath{\mathrm{i}}\xspace}

 \def\Pp      {\ensuremath{\mathrm{p}}\xspace}

 \def\Ps      {\ensuremath{\mathrm{s}}\xspace}

 \def\thebaroffset{0.0em}
}
{
 
 \def\Pgamma      {\ensuremath{\gamma}\xspace}

 \def\Pmu         {\ensuremath{\mu}\xspace}                 
 \def\Pnu         {\ensuremath{\nu}\xspace}                 
                  
 \def\Ppi         {\ensuremath{\pi}\xspace}

 \def\Pphi        {\ensuremath{\phi}\xspace}

 \def\Ppsi        {\ensuremath{\psi}\xspace}                 
                  
 \mathchardef\PDelta="7101
 \mathchardef\PXi="7104
 \mathchardef\PLambda="7103
 \mathchardef\PSigma="7106
 \mathchardef\POmega="710A
 \mathchardef\PUpsilon="7107
                  
 \def\PB      {\ensuremath{B}\xspace}                 
                  
 \def\PD      {\ensuremath{D}\xspace}

 \def\PJ      {\ensuremath{J}\xspace}                 
 \def\PK      {\ensuremath{K}\xspace}

 \def\Pb      {\ensuremath{b}\xspace}                 
 \def\Pc      {\ensuremath{c}\xspace}

 \def\Ph      {\ensuremath{h}\xspace}                 
 \def\Pi      {\ensuremath{i}\xspace}

 \def\Pp      {\ensuremath{p}\xspace}

 \def\Ps      {\ensuremath{s}\xspace}

 \def\thebaroffset{0.18em}
}
\newcommand{\offsetoverline}[2][\thebaroffset]{\kern #1\overline{\kern -#1 #2}}%

\makeatletter
\ifcase \@ptsize \relax
  \newcommand{\miniscule}{\@setfontsize\miniscule{4}{5}}
\or
  \newcommand{\miniscule}{\@setfontsize\miniscule{5}{6}}
\or
  \newcommand{\miniscule}{\@setfontsize\miniscule{5}{6}}
\fi
\makeatother

\DeclareRobustCommand{\optbar}[1]{\shortstack{{\miniscule (\rule[.5ex]{1.25em}{.18mm})}
  \\ [-.7ex] $#1$}}

\def\mup        {{\ensuremath{\Pmu^+}}\xspace}
\def\mun        {{\ensuremath{\Pmu^-}}\xspace} 

\def\mumu       {{\ensuremath{\Pmu^+\Pmu^-}}\xspace}

\def\neu        {{\ensuremath{\Pnu}}\xspace}
\def\neub       {{\ensuremath{\overline{\Pnu}}}\xspace}
\def\neum       {{\ensuremath{\neu_\mu}}\xspace}
\def\neumb      {{\ensuremath{\neub_\mu}}\xspace}

\def\squark    {{\ensuremath{\Ps}}\xspace}

\def\cquark    {{\ensuremath{\Pc}}\xspace}

\def\bquark    {{\ensuremath{\Pb}}\xspace}

\def\pion   {{\ensuremath{\Ppi}}\xspace}

\def\pim    {{\ensuremath{\pion^-}}\xspace}

\def\kaon    {{\ensuremath{\PK}}\xspace}

\def\KorKbar {\kern \thebaroffset\optbar{\kern -\thebaroffset \PK}{}\xspace}

\def\Kp      {{\ensuremath{\kaon^+}}\xspace}
\def\Km      {{\ensuremath{\kaon^-}}\xspace}

\def\D       {{\ensuremath{\PD}}\xspace}

\def\DorDbar {\kern \thebaroffset\optbar{\kern -\thebaroffset \PD}\xspace}

\def\Dp      {{\ensuremath{\D^+}}\xspace}
\def\Dm      {{\ensuremath{\D^-}}\xspace}

\def\DpDm    {\ensuremath{\Dp {\kern -0.16em \Dm}}\xspace}

\def\B       {{\ensuremath{\PB}}\xspace}
\def\Bbar    {{\ensuremath{\offsetoverline{\PB}}}\xspace}

\def\BorBbar {\kern \thebaroffset\optbar{\kern -\thebaroffset \PB}\xspace}

\def\Bd      {{\ensuremath{\B^0}}\xspace}

\def\BdorBdbar {\kern \thebaroffset\optbar{\kern -\thebaroffset \Bd}\xspace}
\def\Bu      {{\ensuremath{\B^+}}\xspace}

\def\Bs      {{\ensuremath{\B^0_\squark}}\xspace}
\def\Bsb     {{\ensuremath{\Bbar{}^0_\squark}}\xspace}
\def\BsorBsbar {\kern \thebaroffset\optbar{\kern -\thebaroffset \Bs}\xspace}
\def\Bc      {{\ensuremath{\B_\cquark^+}}\xspace}

\def\Bds     {{\ensuremath{\B_{(\squark)}^0}}\xspace}

\def\jpsi     {{\ensuremath{{\PJ\mskip -3mu/\mskip -2mu\Ppsi}}}\xspace}

\def\Y#1S{\ensuremath{\PUpsilon{(#1S)}}\xspace}

\def\Lz          {{\ensuremath{\PLambda}}\xspace}

\def\LorLbar     {\kern \thebaroffset\optbar{\kern -\thebaroffset \PLambda}\xspace}

\def\Lb           {{\ensuremath{\Lz^0_\bquark}}\xspace}

\newcommand{\decay}[2]{\ensuremath{#1\!\to #2}\xspace} 

\def\to                 {\ensuremath{\rightarrow}\xspace}

\def\BTohh        {\decay{\B}{\Ph^+ \Ph'^-}}

\def\AT#1     {\ensuremath{A_{\mathrm{T}}^{#1}}\xspace}           

\def\Bsmm     {\decay{\Bs}{\mup\mun}}
\def\Bdmm     {\decay{\Bd}{\mup\mun}}

\def\C#1      {\ensuremath{\mathcal{C}_{#1}}\xspace}                       
\def\Cp#1     {\ensuremath{\mathcal{C}_{#1}^{'}}\xspace}                    
\def\Ceff#1   {\ensuremath{\mathcal{C}_{#1}^{\mathrm{(eff)}}}\xspace}        
\def\Cpeff#1  {\ensuremath{\mathcal{C}_{#1}^{'\mathrm{(eff)}}}\xspace}       
\def\Ope#1    {\ensuremath{\mathcal{O}_{#1}}\xspace}                       
\def\Opep#1   {\ensuremath{\mathcal{O}_{#1}^{'}}\xspace}                    

\newcommand{\aunit}[1]{\ensuremath{\text{\,#1}}}       

\newcommand{\tev}{\aunit{Te\kern -0.1em V}\xspace}
\newcommand{\gev}{\aunit{Ge\kern -0.1em V}\xspace}
\newcommand{\mev}{\aunit{Me\kern -0.1em V}\xspace}
\newcommand{\kev}{\aunit{ke\kern -0.1em V}\xspace}
\newcommand{\ev}{\aunit{e\kern -0.1em V}\xspace}
 
\newcommand{\mevc}{\ensuremath{\aunit{Me\kern -0.1em V\!/}c}\xspace}
\newcommand{\gevc}{\ensuremath{\aunit{Ge\kern -0.1em V\!/}c}\xspace}
\newcommand{\mevcc}{\ensuremath{\aunit{Me\kern -0.1em V\!/}c^2}\xspace}
\newcommand{\gevcc}{\ensuremath{\aunit{Ge\kern -0.1em V\!/}c^2}\xspace}

\def\fb   {\ensuremath{\aunit{fb}}\xspace}
\def\invfb   {\ensuremath{\fb^{-1}}\xspace}

\def\ps   {\ensuremath{\aunit{ps}}\xspace}

\newcommand{\chisqip}{\ensuremath{\chi^2_{\text{IP}}}\xspace}

\def\gsim{{~\raise.15em\hbox{$>$}\kern-.85em
          \lower.35em\hbox{$\sim$}~}\xspace}
\def\lsim{{~\raise.15em\hbox{$<$}\kern-.85em
          \lower.35em\hbox{$\sim$}~}\xspace}

\def\sPlot{\mbox{\em sPlot}\xspace}

\def\pt         {\ensuremath{p_{\mathrm{T}}}\xspace}

\def\photos     {\mbox{\textsc{Photos}}\xspace}

\def\tell1  {TELL1\xspace}
\def\ukl1   {UKL1\xspace}

\def\Bdobslimitnf{\ensuremath{2.6\times 10^{-10}}\xspace} 
\def\Bsmmgobslimitnf{\ensuremath{2.0\times 10^{-9}}\xspace} 

\def\Bsbr{\ensuremath{\left(3.09^{\,+\,0.46\,+\,0.15}_{\,-\,0.43\,-\,0.11}\right)\times 10^{-9}}\xspace}
\def\Bdbr{\ensuremath{\left(1.2^{\,+\,0.8}_{\,-\,0.7}\pm 0.1\right)\times 10^{-10}}\xspace}
\def\Bsmmgbr{\ensuremath{\left(-2.5\pm1.4\pm0.8\right)\times 10^{-9} \mbox{ with } m_{\mu\mu}>4.9\gevcc}\xspace}

\def\Bstau{\ensuremath{2.07 \pm 0.29\pm 0.03 \ps}\xspace}

\def\Bdsigma{\ensuremath{1.7}\xspace}
\def\Bsmmgsigma{\ensuremath{1.6}\xspace}
\def\Bssigma{\ensuremath{10}\xspace}

\def\Bsbrcomb{\ensuremath{\left(2.69 ^{\,+\,0.37}_{\,-\,0.35}\right)\times 10^{-9}}\xspace}
\def\Bdlimcomb{\ensuremath{1.9 \times 10^{-10}}\xspace}

\def\taumumusm{\ensuremath{1.620\pm 0.007 \ps \xspace}}

\newcommand\runone{Run~1\xspace}
\newcommand\runtwo{Run~2\xspace}

\def\BDT{BDT\xspace}
\newcommand{\comment}[1]{}

\newcommand{\BRof}[1]{\ensuremath{{\cal B}(#1)}\xspace}
\newcommand{\BRofmcut}[1]{\ensuremath{{\cal B}(#1;m_{\mu\mu}>4.9\gevcc)}\xspace}

\definecolor{darkred}{rgb}{0.6,0.0,0.0}
\definecolor{darkgreen}{rgb}{0.0,0.5,0.0}
\definecolor{lightgreen}{rgb}{0.75,1.0,0.75}
\definecolor{lightred}{rgb}{1.0,0.75,0.75}
\definecolor{lightblue}{rgb}{0.75,0.75,1.0}
\definecolor{darkblue}{RGB}{100,100,200}
\definecolor{verylightblue}{rgb}{0.9,0.9,1.0}
\definecolor{verylightred}{rgb}{1.0,0.9,0.9}
\definecolor{lightgray}{rgb}{0.9,0.9,0.9}
\definecolor{verylightgray}{rgb}{0.95,0.95,0.95}
\definecolor{darkgray}{rgb}{0.75,0.75,0.75}
\definecolor{orange}{rgb}{1.0,0.75,0.0}

\def\tmumu{\ensuremath{\tau_{\mu^+\mu^-}}\xspace}

\def\ADeltaGamma{\ensuremath{A^{\mu\mu}_{\Delta\Gamma_s}}\xspace}

\newcommand{\CLs}{\ensuremath{\textrm{CL}_{\textrm{s}}}\xspace}

\newcommand{\Bsd}{\Bds}

\newcommand{\Bsmumu}{\decay{\Bs}{\mup \mun}}

\newcommand{\Bdmumu}{\decay{\Bd}{\mup \mun}}
\newcommand{\Bsmumugamma}{\decay{\Bs}{\mup \mun \Pgamma}}

\newcommand{\Bdmumugamma}{\decay{\Bd}{\mup \mun \Pgamma}}
\newcommand{\Bdsmumu}{\decay{\Bds}{\mup \mun}}

\newcommand{\BdKpi}{\decay{\Bd}{\Kp \pim}}

\newcommand{\Bpimumu}{\decay{\B^{0(+)}}{\pi^{0(+)} \mup \mun}}

\newcommand{\BdPiMuNu}{\decay{\Bd}{\pim \mup \neum}}
\newcommand{\BsKMuNu}{\decay{\Bs}{\Km \mup \neum}}

\newcommand{\BuJpsiK}{\decay{\Bu}{\jpsi \Kp}}

\newcommand{\Bhh}{\decay{\Bds}{h^+ h^{\prime -} }}

\newcommand{\BsJpsiPhi}{\decay{\Bs}{\jpsi \Pphi}}

\newcommand{\BsKK}{\decay{\Bs}{\Kp \Km}}

\newcommand{\Lbpmunu}{\decay{\Lb}{\Pp \mun \neumb}} 
\newcommand{\BcJpsiMuNu}{\decay{\Bc}{\jpsi \mup \neum}}

\newcommand{\bsmumugamma}{\Bsmumugamma}

\newcommand{\bsmumug}{\Bsmumugamma}

\newcommand{\bpimumu}{\Bpimumu}

\newcommand{\bujpsik}{\BuJpsiK}

\newcommand{\bbdim}{\ensuremath{b\bar{b}\to \mu^+ \mu^- X}\xspace}

\newcommand{\Bmm}{\Bdsmumu}

\newcommand{\Bmumu}{\Bdsmumu}
\newcommand{\bdkpi}{\BdKpi}

\def\BorBbars    {\kern 0.18em\optbar{\kern -0.18em B}}

\def\Y#1S{\ensuremath{\Upsilon{(#1S)}}\xspace}

\usepackage{cite} 
\usepackage{mciteplus}

\usepackage{overpic}

\begin{document}

\renewcommand{\thefootnote}{\fnsymbol{footnote}}
\setcounter{footnote}{1}


\begin{titlepage}
\pagenumbering{roman}

\vspace*{-1.5cm}
\centerline{\large EUROPEAN ORGANIZATION FOR NUCLEAR RESEARCH (CERN)}
\vspace*{1.5cm}
\noindent
\begin{tabular*}{\linewidth}{lc@{\extracolsep{\fill}}r@{\extracolsep{0pt}}}
\ifthenelse{\boolean{pdflatex}}
{\vspace*{-1.5cm}\mbox{\!\!\!\includegraphics[width=.14\textwidth]{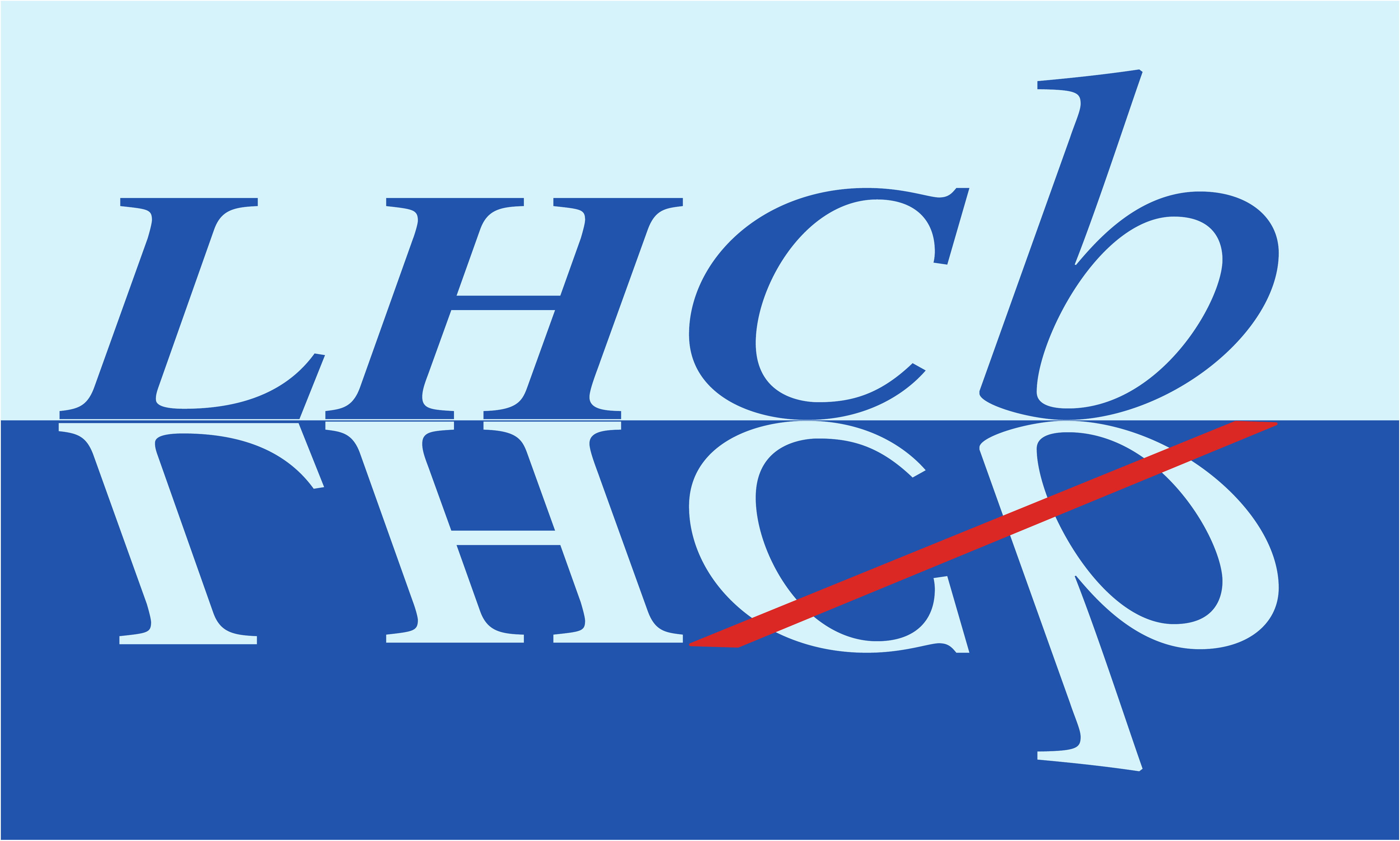}} & &}%
{\vspace*{-1.2cm}\mbox{\!\!\!\includegraphics[width=.12\textwidth]{figs/lhcb-logo.eps}} & &}%
\\
  & & CERN-EP-2021-132 \\  
 & & LHCb-PAPER-2021-007 \\  
 & & 25 February 2022 \\ 
 & & \\
\end{tabular*}

\vspace*{4.0cm}

{\normalfont\bfseries\boldmath\huge
\begin{center}
Analysis of neutral \B-meson decays into two muons
\end{center}
}

\vspace*{2.0cm}

\begin{center}
\paperauthors\footnote{Full author list given at the end of the article.}
\end{center}

\vspace{\fill}


\begin{abstract}
Branching fraction and effective lifetime measurements of the rare decay \Bsmumu and searches for the decays \Bdmumu and \Bsmumugamma are reported using proton-proton collision data collected with the LHCb detector at centre-of-mass energies of  7, 8 and $13\tev$, corresponding to a luminosity of~9\invfb. The branching fraction ${\cal B}(\Bsmumu)=\Bsbr$  and the effective lifetime \mbox{$\tau(\Bsmm)=\Bstau$} are measured, where the first uncertainty is statistical and the second systematic. No significant signal for \Bdmumu and \Bsmumugamma decays is found and upper limits $\mbox{\BRof \Bdmumu} < \Bdobslimitnf$ and \mbox{$\BRof \Bsmumugamma < \Bsmmgobslimitnf$} at the 95\% CL are determined, where the latter is limited to the range \mbox{$m_{\mu\mu} > 4.9 \gevcc$}. 
The results are in agreement with the Standard Model expectations.

\end{abstract}

\vspace*{2.0cm}

\begin{center}
  Published in
  Phys.~Rev.~Lett. 128, (2022) 041801
\end{center}

\vspace{\fill}

{\footnotesize 
\centerline{\copyright~\papercopyright. \href{\paperlicenceurl}{\paperlicence}.}}
\vspace*{2mm}

\end{titlepage}


\newpage
\setcounter{page}{2}
\mbox{~}
%
%
%
%


\renewcommand{\thefootnote}{\arabic{footnote}}
\setcounter{footnote}{0}

\cleardoublepage


\pagestyle{plain} 
\setcounter{page}{1}
\pagenumbering{arabic}



\noindent 
The leptonic decays \Bdmumu and \Bsmumu are very rare in the Standard Model (SM) of particle physics because they only proceed via quantum-loop transitions and are helicity and Cabibbo-Kobayashi-Maskawa (CKM) suppressed. The SM predictions of their time-integrated branching fractions, \mbox{\BRof \Bsmumu = $(3.66 \pm 0.14) \times 10^{-9}$} and \mbox{\BRof \Bdmumu = $(1.03 \pm 0.05) \times 10^{-10}$}~\cite{Bobeth:2013uxa,Beneke:2019slt}, have small uncertainties owing to the leptonic final state and to the progress in lattice QCD calculations~\cite{Aoki:2019cca,Bazavov:2017lyh,Bussone:2016iua,Dowdall:2013tga,Hughes:2017spc}.
Precise measurements of these observables may reveal discrepancies with the expected values due to the existence of new particles contributing to the decay amplitudes, such as heavy $Z^\prime$ gauge bosons, leptoquarks or non-SM Higgs bosons (see e.g.~\cite{Altmannshofer:2017wqy}). For these reasons, over the last decades the measurement of the \Bdmumu and \Bsmumu rates has attracted considerable interest in both theoretical and experimental communities, culminating with the observation of the \Bsmumu decay using the joint LHCb and CMS  Run~1 data sets~\cite{LHCb-PAPER-2014-049} followed by the first single-experiment observation by LHCb~\cite{LHCb-PAPER-2017-001}. Recently, the LHCb measurement has been combined with the ATLAS and CMS measurements~\cite{Aaboud:2018mst,Sirunyan:2019xdu} resulting in $\BRof \Bsmumu = \mbox{\Bsbrcomb}$ and $\BRof \Bdmumu < \mbox{\Bdlimcomb}$ at 95\% confidence level (CL)~\cite{LHCb-CONF-2020-002}, consistent with SM predictions within two standard deviations.

The \Bdmumu and \Bsmumu decays can be accompanied by the emission of final-state radiation (FSR) from the muons or initial-state radiation (ISR) from the valence quarks, with negligible interference between the two processes~\cite{Melikhov:2004mk,Kozachuk:2017mdk,Dettori:2016zff}.
Photons from FSR are predominantly soft and their contribution is included experimentally in the reconstructed \Bsd mass shape as a radiative tail. On the contrary, the ISR process, indicated as \bsmumugamma in this Letter, is characterised by a larger momentum of the photon. This contribution, searched for in the present analysis for the first time, has a SM branching fraction of the order of $10^{-10}$ for a dimuon mass above the lower bound of the search window, 4.9\gevcc, and can be affected by new physics contributions in a different way than the \Bsmumu decay~\cite{Eilam:1996vg,Aliev:1996ud,Geng:2000fs,Melikhov:2004mk,Kozachuk:2017mdk,Dubnicka:2018gqg,Beneke:2020fot,Guadagnoli:2017quo}. Throughout this Letter, \Bmm candidates include \Bsmumu, \Bdmumu or \mbox{\Bsmumugamma} decays with the dimuon pair selected in the mass range [4900, 6000]\mevcc and the photon not reconstructed~\cite{Dettori:2016zff}. The contribution from \Bdmumugamma decays is considered negligible compared to that from \Bsmumugamma because of the additional CKM suppression and the mass shift to lower values.

 The \Bs mass eigenstates are characterised by a sizeable difference in their decay widths ($\Delta\Gamma_s$) compared to their average value ($1/\tau_{B_s}$), such that \mbox{$y_s\equiv \tau_{B_s}\Delta\Gamma_s/2=0.065\pm0.003$}~\cite{HFLAV18}. The effective lifetime, defined as the average decay time, is given by~\cite{DeBruyn:2012wj}
 \begin{equation*}
 \tau_{\mu^+\mu^-}=\frac{\tau_\Bs(1+2\ADeltaGamma y_s+y^2_s)}{(1-y^2_s)(1+\ADeltaGamma y_s)}\,,
 \end{equation*}
where $\ADeltaGamma=1$ ($-1$) if only the heavy (light) \Bs eigenstate can decay to the $\mu^+\mu^-$ final state. In the SM $\ADeltaGamma=1$, but any value in the range $[-1, 1]$ may be possible in new physics scenarios. As a consequence, the effective lifetime of \Bsmumu decays can probe new physics in a way complementary to the branching fraction~\cite{DeBruyn:2012wk}. 

This Letter reports improved measurements of the \Bsmm and \Bdmm time-integrated branching fractions and of the \Bsmm effective lifetime, which supersede the previous LHCb results~\cite{LHCb-PAPER-2017-001}, and a first search for \bsmumug decays. A more comprehensive description of these measurements is reported in a companion article~\cite{LHCb-PAPER-2021-008}.
Inclusion of charge-conjugated processes is implied throughout the Letter. 
Results are based on data collected with the LHCb detector in the years 2011-2012 and 2015-2018, corresponding to an integrated luminosity of 1\invfb  of proton-proton ($pp$) collisions at a centre-of-mass energy $\sqrt{s}=7\tev$, 2\invfb at $\sqrt{s}=8\tev$ and 6\invfb recorded at $\sqrt{s}=13\tev$. 
The first two data sets are referred to as Run 1 and the latter as Run 2.

The \lhcb detector is a single-arm forward spectrometer covering the pseudorapidity range \mbox{$2<\eta<5$}, described in detail in Refs.~\cite{Alves:2008zz,LHCb-DP-2014-002}. 
The simulated events used in this analysis are produced with the software described in Refs.~\cite{Sjostrand:2007gs,*Sjostrand:2006za,LHCb-PROC-2010-056,Lange:2001uf,Allison:2006ve,*Agostinelli:2002hh,LHCb-PROC-2011-006} taking into account the variations of the accelerator and detector conditions over time. In particular, FSR is simulated using \photos~\cite{davidson2015photos}. ISR \bsmumugamma decays are simulated according to the study in Ref.~\cite{Melikhov:2004mk}.
The analysis strategy is similar to that employed in Ref.~\cite{LHCb-PAPER-2017-001}, optimised to enhance the sensitivity to both \Bs\ and \Bd\ decays to \mumu.
After loose trigger and selection requirements, \Bmm candidates are classified based on the dimuon mass and the output variable, BDT, of a boosted decision tree classifier~\cite{Breiman,Adaboost} designed to distinguish signal from combinatorial background. To avoid the experimenter's bias, the candidates in the region [5200, 5445]\mevcc, where the \Bsmumu and \Bdmumu signal processes peak, were not examined until the full procedure had been finalised.
The signal yields are determined from a maximum-likelihood fit to the dimuon mass distribution of the candidates in regions of BDT, and are converted into branching fractions using the decays \bdkpi and \bujpsik, with $\jpsi\to \mu^+ \mu^-$, as normalisation modes. These decays have been chosen for their relatively large and well-measured branching fractions and because they share the same topology or a dimuon pair in the final state with the signal.
The \Bsmumu effective lifetime is measured from the background-subtracted decay-time distribution of signal candidates.

Events are selected by a hardware trigger followed by a software trigger~\cite{Aaij:2012me}. The \Bmm candidates are predominantly selected by single-muon and dimuon triggers. 
The \BuJpsiK candidates are selected in the same way except for  a different dimuon mass requirement in the software trigger. Candidate \Bhh decays, with $h^{(\prime)}=\pi$ or~$K$, are used as control and normalisation channels and are triggered independently of the \Bds decay products to avoid selection biases.

The \Bmm candidates are reconstructed by combining two oppositely charged tracks with transverse momentum with respect to the proton beam direction, $p_{\rm T}$, in the range \mbox{$0.25<p_{\rm T}<40\gevc$}, momentum $p< 500\gevc$, and high-quality muon identification~\cite{Archilli:2013npa}. 
The muon candidates are required to be inconsistent with originating from any primary $pp$ interaction vertex (PV) and to form a good quality secondary vertex well displaced from any PV.
In the selection, \Bsd candidates must have a decay time less than 13.25\ps, $p_{\rm T} > 0.5 \gevc$ and they must be consistent with originating from at least one PV. 
A \Bsd candidate is rejected if either of the two candidate muons combined with any other oppositely charged muon candidate in the event has a mass consistent with the $J/\psi$ mass~\cite{jpsimass}.
The \Bhh selection is the same as that of \Bmumu, except that the muon identification criteria are replaced with hadron identification requirements and the \jpsi veto is not applied. The \BuJpsiK decay is reconstructed by combining a muon pair, consistent with a \jpsi from a detached vertex, and a kaon candidate inconsistent with originating from any PV in the event. The selection criteria for signal and normalisation candidates include a loose requirement on the response of a different multivariate classifier, described in Refs.~\cite{LHCb-PAPER-2012-007,LHCb-PAPER-2021-008}.

\begin{figure}[t!]
\begin{center}
\includegraphics*[width=.9\columnwidth]{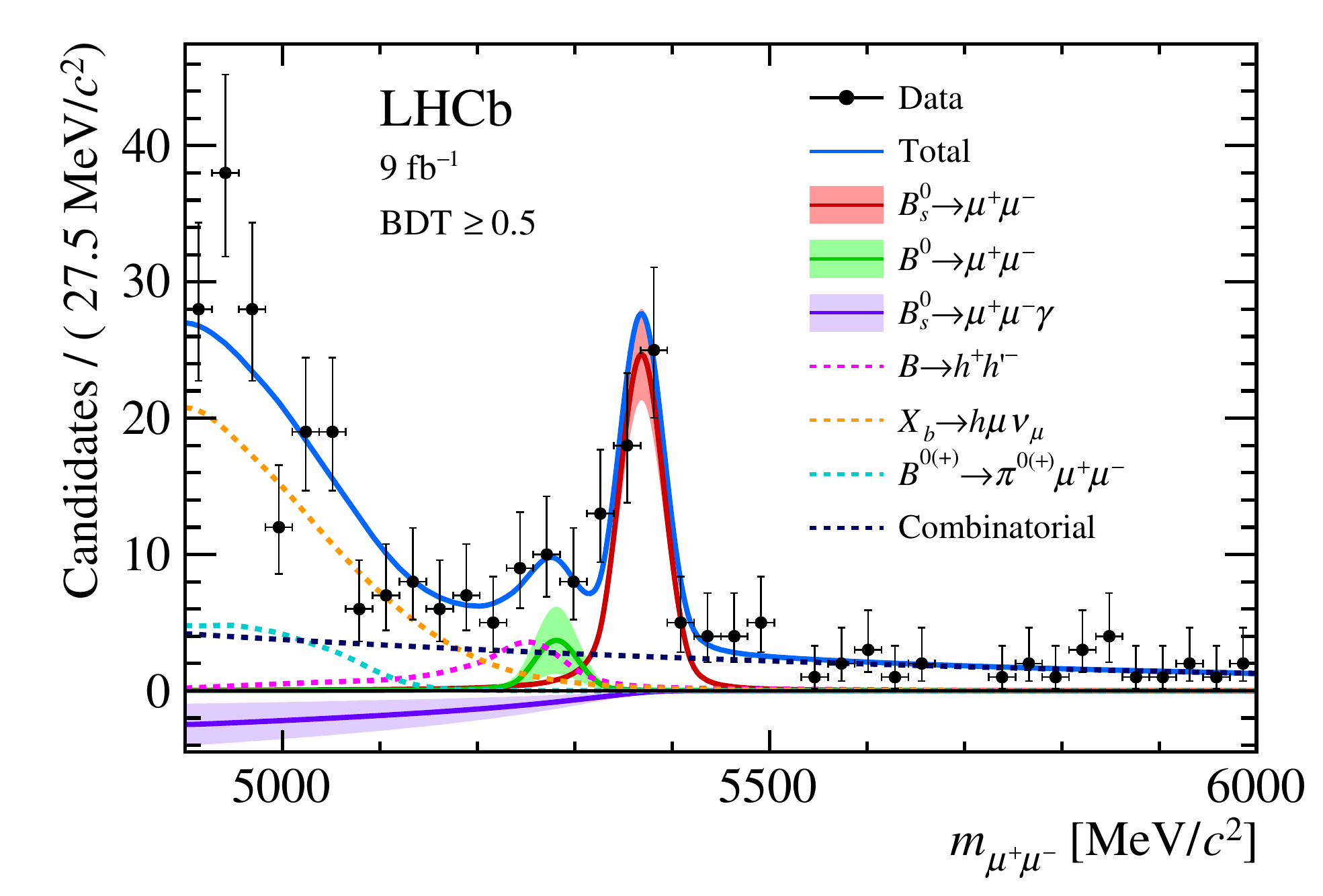}
\end{center}
\caption{\small Mass distribution of the selected \Bmm candidates (black dots) with ${\rm \BDT}>0.5$. The result of the fit is overlaid and the different components are detailed: \Bsmumu (red solid line), \Bdmumu (green solid line), \Bsmumugamma (violet solid line), combinatorial background (blue dashed line), \Bhh (magenta dashed line), \BdPiMuNu, \BsKMuNu, \BcJpsiMuNu and \Lbpmunu (orange dashed line), and \bpimumu (cyan dashed line). The solid bands around the signal shapes represent the variation of the branching fractions by their total uncertainty.  
}\label{fig:mass}
\end{figure}

The selected events are dominated by combinatorial background, mainly composed of muons originating from two semileptonic $b$-hadron decays. The separation between signal and combinatorial background is achieved by means of the BDT classifier, which is optimised using simulated samples of \Bsmm events for signal and of \bbdim events for background. The classifier combines information from the following input variables: $\sqrt{\Delta\phi^2+\Delta\eta^2}$, where $\Delta\phi$ and $\Delta\eta$ are the azimuthal angle and pseudorapidity differences between the two muon candidates; the minimum $\chi^2_{\rm IP}$ of the two muons candidates with respect to the PV$_B$, where PV$_B$ is the PV most compatible with the \Bsd candidate trajectory and \chisqip is defined as the difference between the vertex-fit $\chi^2$ of the PV formed with and without the particle in question; the angle between the direction of the \Bsd candidate momentum and the vector joining the \Bsd decay vertex and PV$_B$; the \Bsd candidate vertex-fit $\chi^2$ and impact parameter significance with respect to the PV$_B$; and two isolation variables that quantify how much the other tracks of the event are likely to originate from the same hadron decay as the signal tracks. 
The \BDT variable is constructed to be approximately uniform in the range [0,1] for signal, and to peak strongly at zero for background. Its linear correlation with the dimuon mass is below 5\%. 
The Run~1 and Run 2 data sets are each divided into six subsets based on BDT regions with boundaries $0.0$, $0.25$, $0.4$, $0.5$, $0.6$, $0.7$ and $1.0$; candidates having ${\rm BDT}<0.25$ are not included in the fit to the dimuon mass distribution. The  mass distribution of the \mbox{\Bmm} candidates with ${\rm \BDT}>0.5$ is shown in Fig.~\ref{fig:mass}.

The BDT distributions of \Bdsmumu decays are calibrated using simulated samples which have been reweighted to improve the agreement with the data. The $\pt$, $\eta$ and \chisqip quantities of simulated \Bd and \Bs samples are corrected~\cite{Rogozhnikov:2016bdp} 
using data samples of \BuJpsiK and \BsJpsiPhi decays, respectively. The event occupancy is also corrected, separately for each BDT region, by comparing the fraction of \BuJpsiK candidates in four intervals of the number of tracks in simulated events and in data.
To align the reconstruction with that of the \Bsmumu signal, the BDT response for the \BuJpsiK candidates is evaluated using the information from the final state muons and the $B^+$ candidate, with two exceptions: the $B$ vertex-fit $\chi^2$ is replaced with that of the $\jpsi$, and the muon isolation variables are computed without considering the final-state kaon. The effect of the trigger selection on the BDT distribution is estimated using control channels in data. 
The resulting \Bdmm and \Bsmm \BDT variable distributions are found to be compatible with that of \mbox{\BdKpi} decays selected in data when corrected for the different trigger and particle identification selection and, in the case of \Bsmm, the different lifetime. 

The mass distributions of the \Bsmm and \Bdmm signals are described by two-sided Crystal Ball functions~\cite{Skwarnicki:1986xj} with core Gaussian parameters calibrated from the mass distributions of \BsKK and \BdKpi data samples, respectively. A mass resolution of about 22\mevcc is determined by interpolating the measured resolutions of charmonium and bottomonium resonances decaying into two muons. The radiative tails are obtained from simulation~\cite{Golonka:2005pn}. Small differences in the resolution and tail parameters of the mass shape for the different BDT regions are taken into account. The mass distribution of the \Bsmumugamma decays is described with a threshold function modelled on simulated events that were generated using the theoretical predictions of Refs.~\cite{Melikhov:2004mk,Kozachuk:2017mdk}, convoluted with the experimental resolution.

The signal branching fractions are determined using the relation
\begin{equation*}
\BRof \Bdsmumu=\frac{{\cal B}_{\rm norm}  \,{\rm \epsilon_{\rm norm}}\,f_{\rm norm} }{ N_{\rm norm}\,{\rm \epsilon_{sig}} \,f_{d(s)} } \times N_{\Bmumu}\equiv\alpha^{\rm norm}_{\Bdsmumu} \times N_{\Bdsmumu},
\end{equation*}
where $N_{\Bmumu}$ is the signal yield determined in the mass fit, $N_{\rm norm}$ is the number of selected normalisation decays (\BuJpsiK or \BdKpi), ${\cal B}_{\rm norm}$ the corresponding branching fraction~\cite{PDG2020}, and ${\rm \epsilon_{sig}}$ (${\rm \epsilon_{\rm norm}}$) is the total efficiency for the signal (normalisation) channel. For each signal mode, the two single event sensitivities, $\alpha^{\rm norm}_{\Bdsmumu}$, are then averaged in a combined $\alpha_{\Bdsmumu}$ taking the correlations into account. 
The fraction $f_{d(s)}$ indicates the probability for a $b$ quark to fragment into a $B^0_{(s)}$ meson. The value of $f_s/f_d$ has been measured by LHCb to be $0.254 \pm 0.008$ in $pp$ collision data at $\sqrt{s}=13\tev$, while the average value in Run~1 is lower by a factor of $1.064\pm 0.007$~\cite{LHCb-PAPER-2020-046}. The fragmentation probabilities for the $B^0$ and $B^+$ are assumed to be equal, hence $f_{\rm norm}=f_d$ for both normalisation modes.

 The acceptance, reconstruction and selection efficiencies are computed with samples of simulated events generated with the decay-time distribution predicted by the SM. The tracking and particle identification efficiencies are determined using control channels in data~\cite{LHCb-DP-2013-002,PIDCalib}. 
The trigger efficiencies are evaluated with control channels in data~\cite{Tolk:1701134}.

The yields of selected \BuJpsiK and \bdkpi decays are \mbox{$(4733\pm 3)\times 10^3$} and \mbox{$(94\pm 1)\times 10^3$}, respectively. The normalisation factors measured with the two channels are consistent and their weighted averages,  taking correlations into account, are $\alpha_{\Bsmumu}=(3.51\pm 0.13)\times 10^{-11}$, $\alpha_{\Bdmumu}=(9.20\pm 0.17)\times 10^{-12}$ and \mbox{$\alpha_{\Bsmumugamma}=(4.57\pm 0.17)\times 10^{-11}$}. Assuming SM predictions for the branching fractions, the analysed data sample is expected to contain an average of $104\pm 6$ \Bsmm, $11\pm 1$ \Bdmm and about 2 \Bsmumugamma decays in the $\rm{BDT}>0.25$ range and in the mass range [4900, 6000]\mevcc.

The combinatorial background is distributed exponentially over the whole mass range. In addition, the \Bd and \Bs signal regions and the low-mass sideband are populated by background from specific $b$-hadron decays divided into two categories: those with the misidentification of at least one hadron as a muon and those where two real muons are present and the decay is partially reconstructed. 
The first category includes \Bhh, \BdPiMuNu, $B^0_s \to K^- \mu^+ \nu_{\mu}$, and \Lbpmunu decays, of which branching fractions are taken from Refs.~\cite{PDG2020,LHCb-PAPER-2020-038,LHCb-PAPER-2015-013}. The mass and BDT distributions of these decays are determined from simulated samples after calibrating the $K\to\mu$, $\pi\to\mu$ and $p\to\mu$ momentum-dependent misidentification probabilities using control channels in data. 
An independent estimate of the \Bhh background yield is obtained by extracting the yields of misidentified \Bhh decays from the mass spectrum of $\pi^+\mu^-$ or $K^+\mu^-$ combinations in data, and rescaling the observed yields according to the misidentification probabilities.
The difference with respect to the result from the first method is assigned as a systematic uncertainty.
The second category of background in the low-mass sideband includes the decays \mbox{$B^+_c \to J/\psi \mu^+ \nu_{\mu}$}, with $J/\psi  \to \mu^+ \mu^-$, and \mbox{\bpimumu}, which have at least two muons in the final state. The rate of \mbox{$B^+_c \to J/\psi \mu^+ \nu_{\mu}$} decays is evaluated from Refs.~\cite{LHCb-PAPER-2014-050,LHCb-PAPER-2014-025} and those of \mbox{\bpimumu} decays from Refs.~\cite{LHCb-PAPER-2015-035,Wang:2012ab}. 
The expected yields of the background contributions originating from specific processes are estimated by normalising to the \BuJpsiK decay, except for the \Bhh decays, which are normalised to the \BdKpi channel. 
Their expected yields with ${\rm \BDT}>0.25$ in the full mass range are $37\pm 2$ \Bhh, $161\pm 6$ \BdPiMuNu, $31\pm 3$ \BsKMuNu, $53\pm 4$ \Bpimumu, $7\pm 3$ \Lbpmunu and $28\pm 1$ \BcJpsiMuNu decays.

The \Bsmm, \Bdmm and \Bsmumugamma branching fractions are determined with a simultaneous unbinned maximum-likelihood fit~\cite{roofit} to the dimuon mass distribution in the BDT regions of the \runone and \runtwo data sets, with $\rm{BDT}>0.25$. 
 The fractions of \Bmumu yield in each BDT region and the parameters of the Crystal Ball functions~\cite{Skwarnicki:1986xj} describing the shapes of the mass distribution are Gaussian constrained according to their expected values and uncertainties. 
The combinatorial background in each BDT region is described by an exponential function with the yield and slope allowed to vary freely, but the slope parameter is common to all regions within a given data set. Each other background is included as a separate component in the fit. Their yields as well as the fractions in each BDT region are Gaussian-constrained according to their expected values, while their mass shapes are determined from simulation and fixed in the fit, separately in each BDT region. Figure~\ref{fig:mass} shows the fit results projected on the dimuon mass distribution for $\mbox{BDT}>0.5$.

The branching fractions of the \mbox{\Bsmm}, \mbox{\Bdmm} and \mbox{\Bsmumugamma} decays obtained from the fit are 
\begin{eqnarray}
&&\BRof \Bsmumu=\Bsbr\,, \nonumber\\
&&\BRof \Bdmumu=\Bdbr\,, \nonumber\\
&&\BRof \Bsmumugamma=\Bsmmgbr\, .\nonumber
\end{eqnarray}
The statistical uncertainty is obtained by re-running the fit with all nuisance parameters fixed to the values found in the default fit. The systematic uncertainties of $\BRof \Bsmumu$  and $\BRof\Bdmumu$ are dominated by the uncertainty on $f_s/f_d$ (3\%) and the knowledge of the background from specific processes  (9\%), respectively.
The correlation between the \Bdmm and \Bsmumu branching fractions is $-11\%$ while that between the \Bsmumugamma and \Bdmm (\Bsmm) branching fractions is $-25\%$ (9\%).

\begin{figure}[t!]
\begin{center}
     \includegraphics*[width=0.95\linewidth]{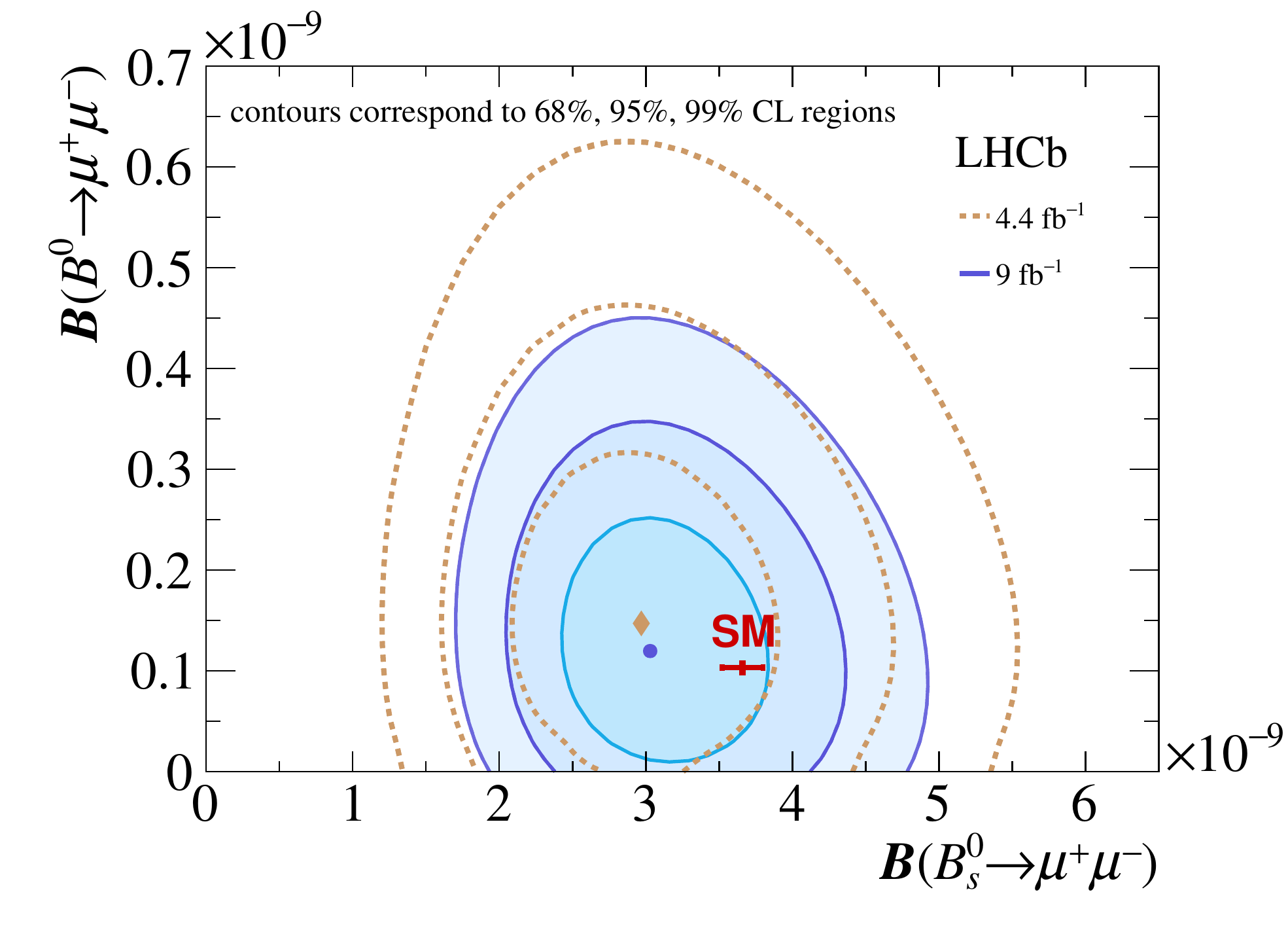}     
     \includegraphics*[width=0.49\linewidth]{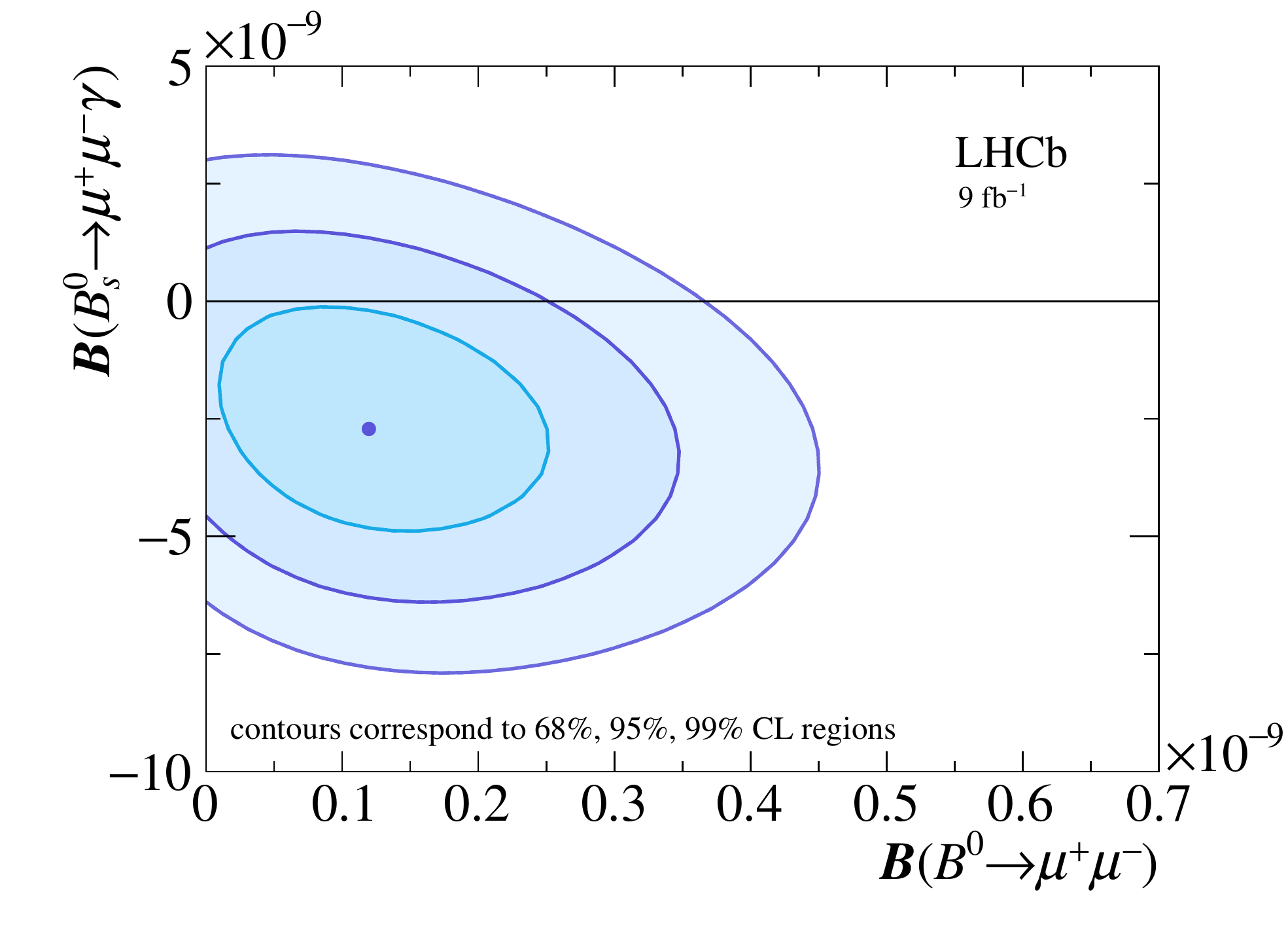}
     \includegraphics*[width=0.49\linewidth]{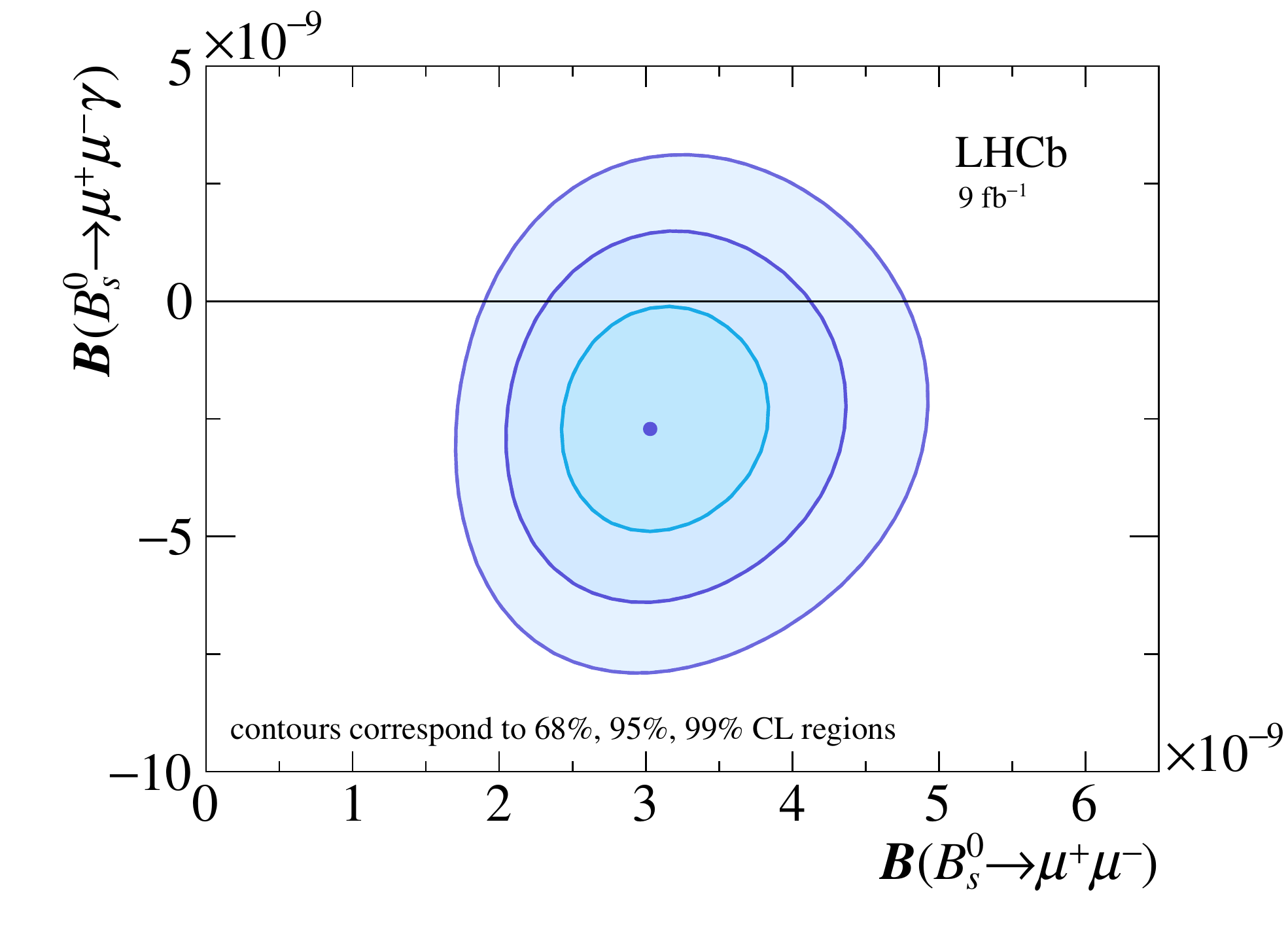}
\end{center}
\caption{\small Two-dimensional profile likelihood of the branching fractions for the decays (top) \mbox{\Bsmm} and \Bdmm, (bottom left) \Bdmm and \Bsmumugamma and (bottom right) \Bsmm and \Bsmumugamma. The \Bsmumugamma branching fraction is limited to the range $m_{\mu\mu}>4.9\gevcc$. The measured central values of the branching fractions are indicated with a blue dot. The profile likelihood contours for 68\%, 95\% and 99\% CL regions of the result are shown as blue contours, while in the top plot the brown contours indicate the previous measurement~\cite{LHCb-PAPER-2017-001} and the red cross shows the SM prediction.}\label{fig:likelihoods}
\end{figure}

Two-dimensional profile likelihoods are evaluated by taking the ratio of the likelihood value of a fit where the parameters of interest are fixed and the likelihood value of the standard fit. They are shown in Figure~\ref{fig:likelihoods} for the possible combinations of two branching fractions.

An excess of \Bsmumu decays with respect to the expectation from background is observed with a significance of about \Bssigma standard deviations ($\sigma$), while the significance of the \Bdmumu signal is $\Bdsigma\sigma$, as determined using Wilks' theorem~\cite{Wilks:1938dza} from the difference in likelihood between fits with and without the specific signal component. The negative fluctuation of the \Bsmumugamma signal has a $\Bsmmgsigma\sigma$ significance.

Since the \Bdmm and \Bsmumugamma signals are not significant, an upper limit on each branching fraction is set using the \CLs method~\cite{Read:2002hq} with a profile likelihood ratio as a one-sided test statistic~\cite{LHCb-PAPER-2016-032}. The likelihoods are computed with the nuisance parameters Gaussian-constrained to their fit values. The test statistic is then evaluated on an ensemble of pseudoexperiments where the nuisance parameters are floated according to their uncertainties.
The resulting upper limit on \BRof \Bdmumu is \mbox{$\Bdobslimitnf$} at 95\%~CL, obtained without constraining the \Bsmumugamma yield.
Similarly, the upper limit on \BRof\Bsmumugamma with $m_{\mu\mu} > 4.9\gevcc$ is evaluated to be \mbox{$\Bsmmgobslimitnf$} at 95\% CL.
Fixing the \Bsmumugamma signal to zero, the \Bsmumu branching fraction increases by about 2\% and the upper limit on $\BRof \Bdmumu$ decreases by about $10\%$.

The selection efficiency of \Bsmm decays depends on the lifetime, introducing a model dependence in the measured time-integrated branching fraction. In the fit the SM value for \tmumu, \taumumusm~\cite{PDG2020}, is assumed, corresponding to $\ADeltaGamma=1$. The model dependence is evaluated by repeating the fit under the assumptions $\ADeltaGamma=0$ and $-1$, finding an increase of the branching fraction with respect to the SM hypothesis of 4.7\% and 10.9\%, respectively. The dependence is approximately linear in the physically allowed \ADeltaGamma range. A similar dependence is present for the \Bsmumugamma decay with a negligible impact on the branching fraction limit.

The criteria used to select data for the \Bsmumu lifetime measurement differ slightly from those used in the branching fraction measurement. As shown in Fig.~\ref{fig:mass}, the contribution from the misidentified background is negligible under the peak, and therefore a narrower dimuon mass range of $[5320,6000]\mevcc$ is selected, while particle-identification requirements are relaxed slightly due to the lower expected contamination from the misidentified background in the \Bsmm signal region, with a corresponding increase in signal efficiency. Finally, candidate \Bsmumu decays are required to fall into two trigger categories: the trigger requirements must be satisfied entirely either by the \Bsmumu candidates themselves, or by objects from the $pp$ collision that do not form part of the \Bsmumu candidate. These more restrictive trigger requirements are imposed in order to improve the modelling of the decay-time dependence of the trigger efficiency in simulation.

In order to determine the \Bsmm effective lifetime the data are divided into two BDT regions [0.35, 0.55] and [0.55, 1.00], with boundaries optimised to achieve the best precision. Fits are performed to the dimuon mass distribution in each BDT region in order to extract background-subtracted decay time distributions using the \sPlot technique~\cite{Pivk:2004ty}. The mass fits used in the background subtraction include \Bsmm and combinatorial background components, where the signal is modelled with the same function as in the branching fraction analysis and the background with exponential functions, with freely-floating slope parameters in each BDT region. The correlation between the reconstructed mass and the reconstructed decay time of the selected candidates is consistent with zero in both data and simulation, as required by the \sPlot technique. 

A simultaneous fit is then performed to the two background-subtracted decay-time distributions, where each distribution is modelled by a single exponential multiplied by an acceptance function that models the decay time dependence of the reconstruction and selection efficiency. The acceptance functions are determined in each BDT region by fitting parametric functions to the efficiency distributions of simulated \Bsmm decays that have been weighted in order to improve the agreement with the data. The correction for the acceptance is validated by measuring the lifetimes of \BdKpi and \BsKK decays in data. The resulting values are $1.510 \pm 0.015\ps$ and $1.435 \pm 0.026\ps$, respectively, where uncertainties are statistical only. These are consistent with the world averages~\cite{PDG2020}. The statistical uncertainty on the measured \BsKK lifetime is taken as the systematic uncertainty associated with the use of simulated events to determine the \Bsmm acceptance function. 

A number of sources of systematic bias are evaluated using a large number of simulated pseudoexperiments. The fit procedure is found to produce an unbiased estimate of the lifetime with uncertainties that provide the correct coverage. The effect of the contamination from \Bdmm, \BTohh and semileptonic $b$-hadron decays in the mass fit is found to introduce a small bias of up to 0.012\ps. The effect of the acceptance on the relative admixture of light and heavy mass eigenstates in the decay-time distribution is found to be negligible. Likewise, the uncertainty in the decay-time distribution of the combinatorial background, the production asymmetry between \Bs and \Bsb mesons and the mismodelling of the acceptance function in simulation is found to have a small effect on the final result. Together, these sources result in a systematic uncertainty of 0.031\ps, which is dominated by the uncertainty on the measured \BsKK lifetime.

\begin{figure}[t]
    \centering
     \includegraphics[width=0.48\linewidth]{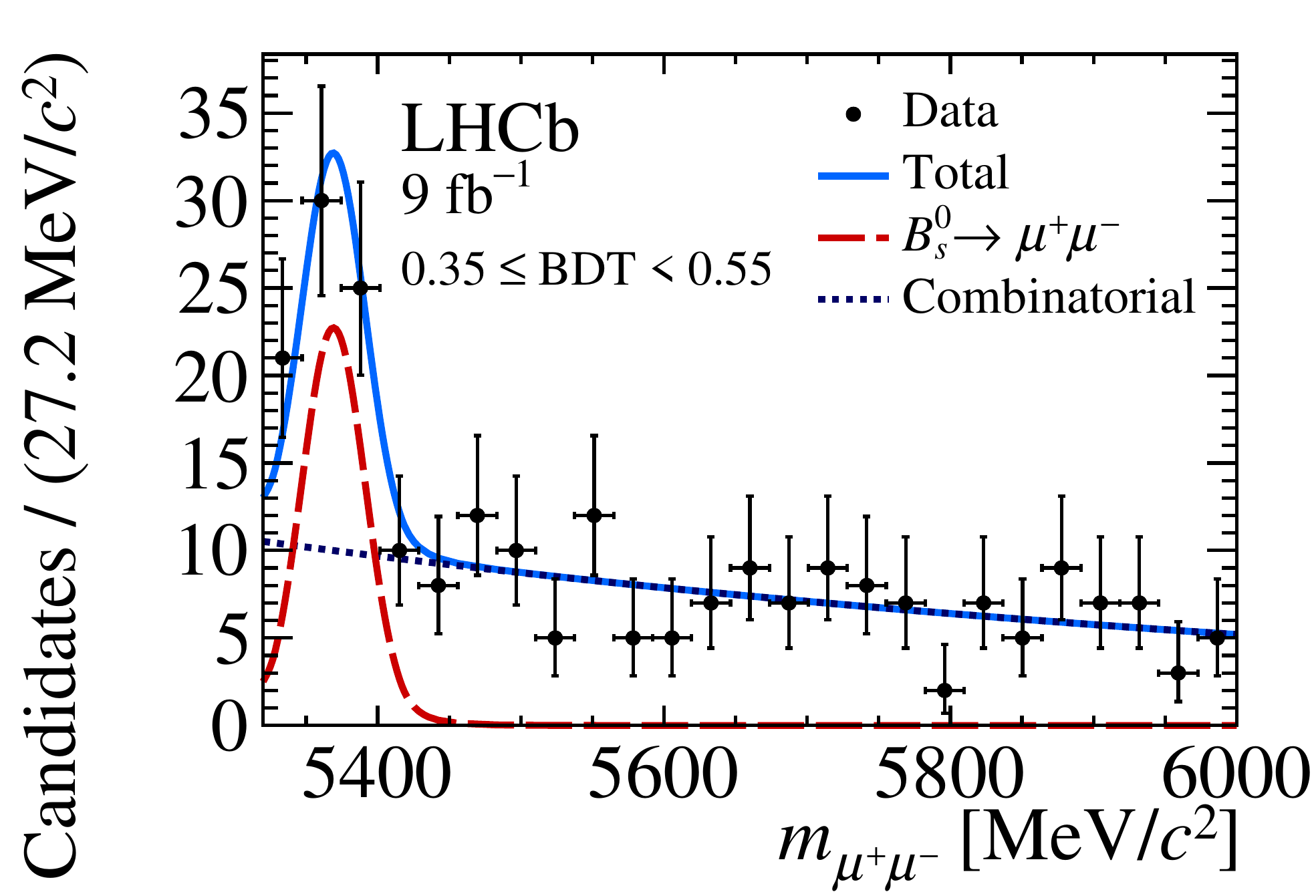}
     \includegraphics[width=0.48\linewidth]{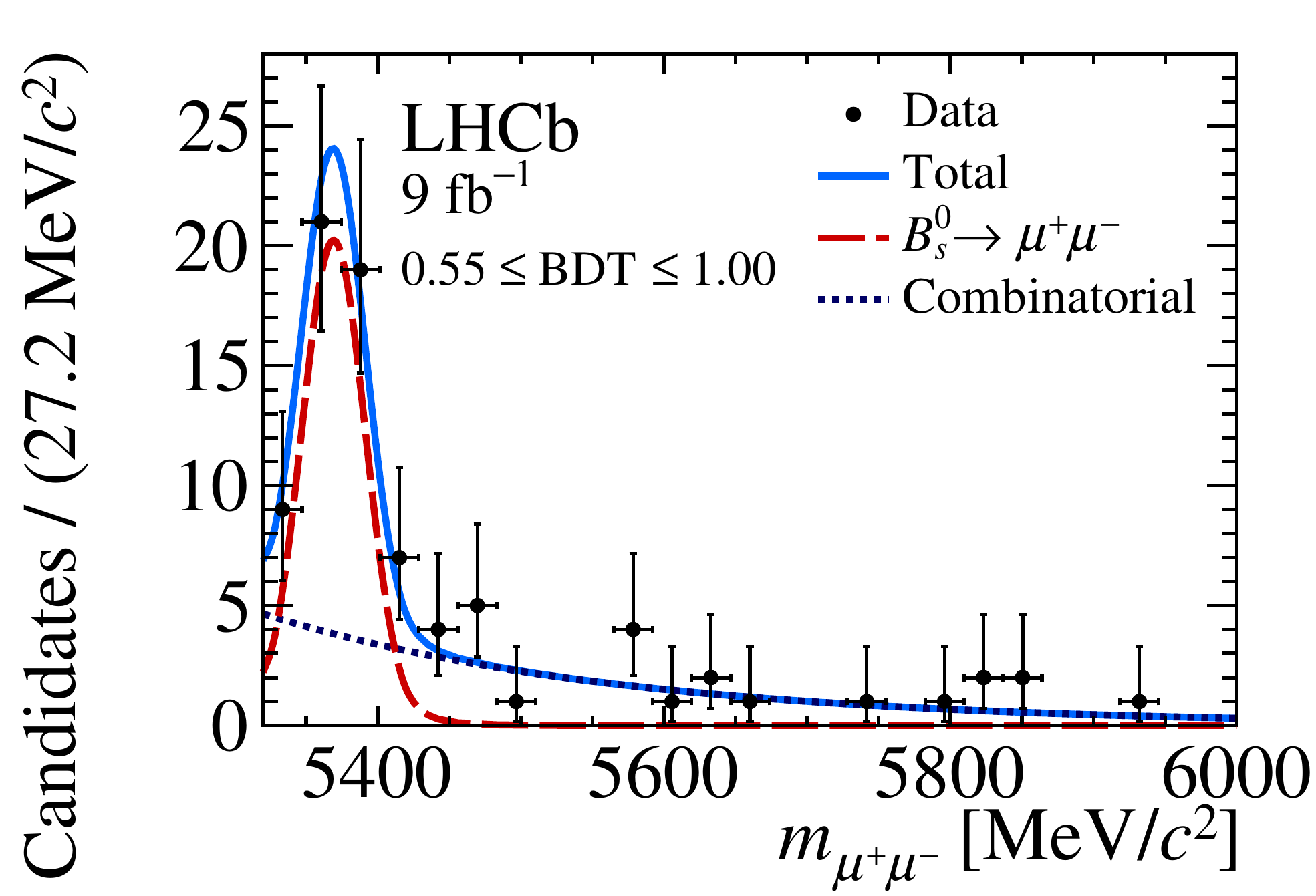}
     \includegraphics[width=0.48\linewidth]{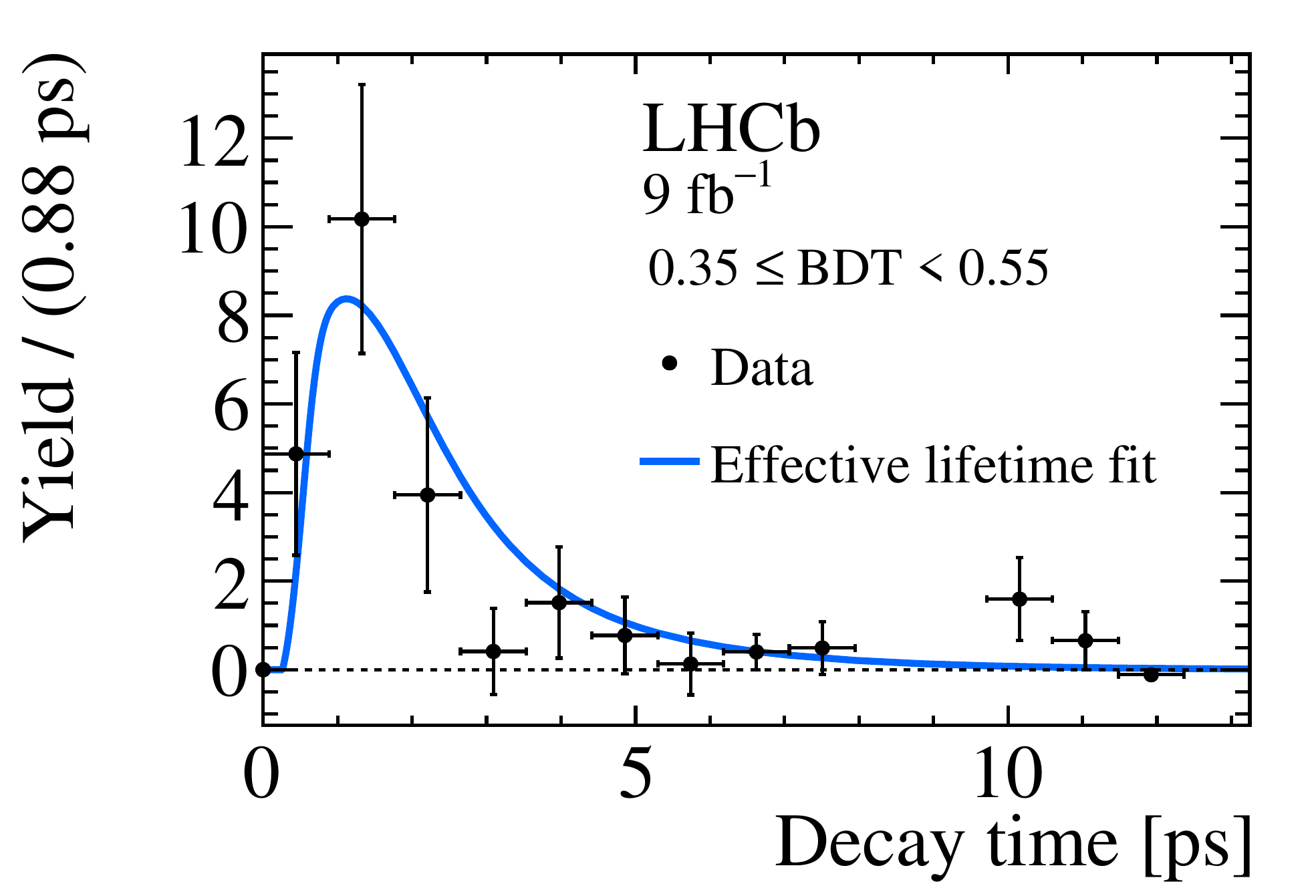}
     \includegraphics[width=0.48\linewidth]{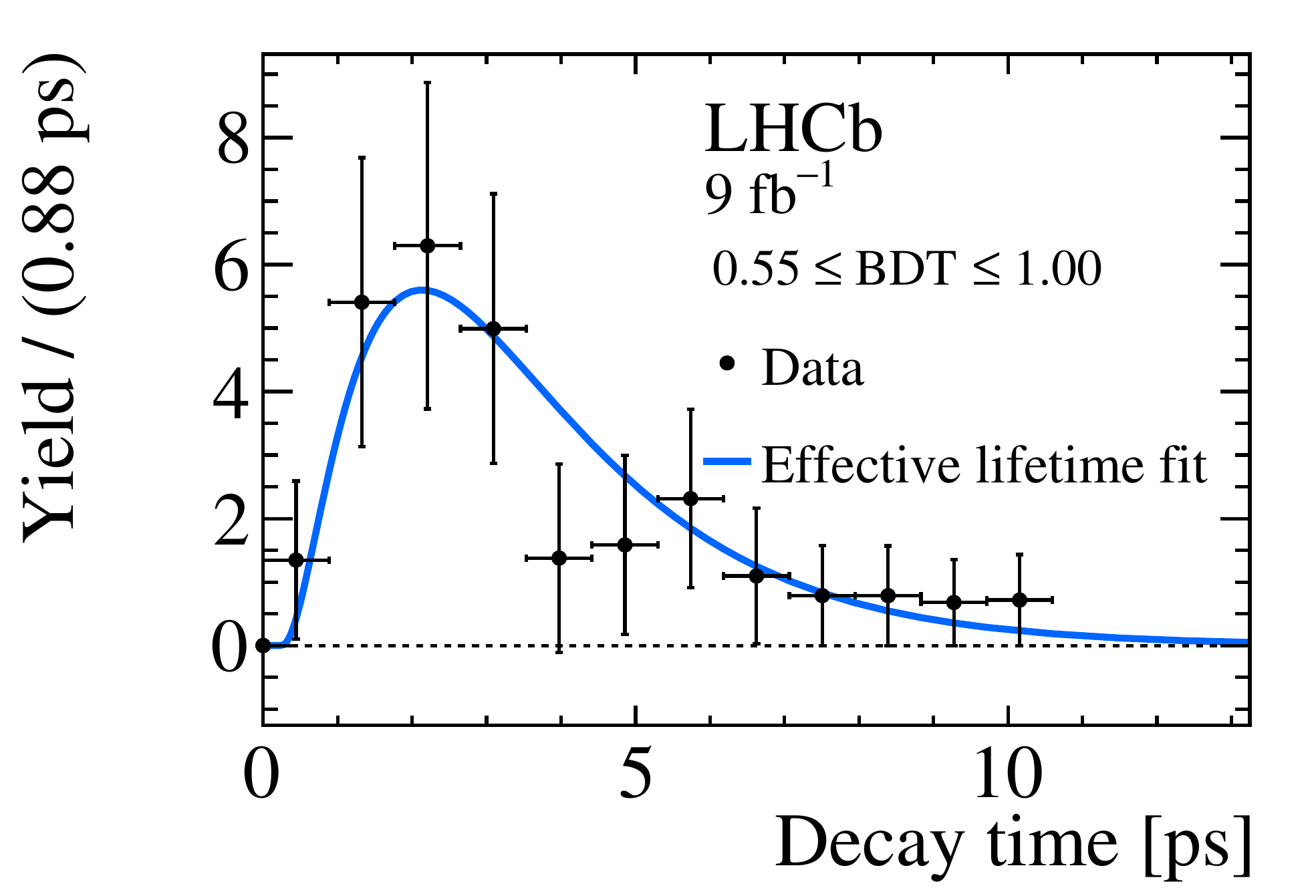}
\caption{Top: dimuon mass distributions with the fit models used to perform the background subtraction superimposed. Bottom: the background-subtracted decay-time distributions with the fit model used to determine the \Bsmm effective lifetime superimposed. The distributions in the low and high BDT regions are shown in the left and right columns, respectively.}\label{fig:lifetime_fit}
\end{figure}

The mass distributions of the selected \Bsmm candidates are shown in \mbox{Fig.~\ref{fig:lifetime_fit} (top) for the two BDT regions}. Figure~\ref{fig:lifetime_fit} (bottom) shows the corresponding background-subtracted \Bsmm decay-time distribution with the fit function superimposed~\cite{roofit}. The effective lifetime is found to be \Bstau, where the first uncertainty is statistical and the second systematic. This value lies outside the range between the lifetimes of the light ($A_{\Delta\Gamma} = -1$) and heavy ($A_{\Delta\Gamma} = +1$) mass eigenstates, which are $\tau_{L} = 1.423 \pm 0.005\ps$ and $\tau_{H} = 1.620 \pm 0.007\ps$ \cite{PDG2020}, but is consistent with these values at 2.2 and 1.5 standard deviations, respectively.

In summary, a new measurement of the rare decay \Bsmm and a search for \Bdmumu and \Bsmumugamma decays has been performed using the full dataset collected by the LHCb experiment during \runone and \runtwo, corresponding to a total integrated luminosity of 9\invfb. 
The time-integrated branching fraction of \Bsmumu is measured to be \mbox{\Bsbr}.
The \Bsmumu effective lifetime is \Bstau. No evidence for \Bdmumu or \Bsmumugamma signals is found, and the upper limits \mbox{$\BRof \Bdmumu < \Bdobslimitnf$} and \mbox{$\BRof \Bsmumugamma < \Bsmmgobslimitnf$} at 95\% CL are set, where the latter is limited to the range \mbox{$m_{\mu\mu} > 4.9 \gevcc$}. The results are in agreement with the SM predictions and can be used to further constrain possible new physics contributions to these observables. 

\section*{Acknowledgements}
%
%
\noindent We express our gratitude to our colleagues in the CERN
accelerator departments for the excellent performance of the LHC. We
thank the technical and administrative staff at the LHCb
institutes.
We acknowledge support from CERN and from the national agencies:
CAPES, CNPq, FAPERJ and FINEP (Brazil); 
MOST and NSFC (China); 
CNRS/IN2P3 (France); 
BMBF, DFG and MPG (Germany); 
INFN (Italy); 
NWO (Netherlands); 
MNiSW and NCN (Poland); 
MEN/IFA (Romania); 
MSHE (Russia); 
MICINN (Spain); 
SNSF and SER (Switzerland); 
NASU (Ukraine); 
STFC (United Kingdom); 
DOE NP and NSF (USA).
We acknowledge the computing resources that are provided by CERN, IN2P3
(France), KIT and DESY (Germany), INFN (Italy), SURF (Netherlands),
PIC (Spain), GridPP (United Kingdom), RRCKI and Yandex
LLC (Russia), CSCS (Switzerland), IFIN-HH (Romania), CBPF (Brazil),
PL-GRID (Poland) and NERSC (USA).
We are indebted to the communities behind the multiple open-source
software packages on which we depend.
Individual groups or members have received support from
ARC and ARDC (Australia);
AvH Foundation (Germany);
EPLANET, Marie Sk\l{}odowska-Curie Actions and ERC (European Union);
A*MIDEX, ANR, IPhU and Labex P2IO, and R\'{e}gion Auvergne-Rh\^{o}ne-Alpes (France);
Key Research Program of Frontier Sciences of CAS, CAS PIFI, CAS CCEPP, 
Fundamental Research Funds for the Central Universities, 
and Sci. \& Tech. Program of Guangzhou (China);
RFBR, RSF and Yandex LLC (Russia);
GVA, XuntaGal and GENCAT (Spain);
the Leverhulme Trust, the Royal Society
 and UKRI (United Kingdom).


\newpage

\addcontentsline{toc}{section}{References}
\setboolean{inbibliography}{true}
\bibliographystyle{LHCb}
\bibliography{main,standard,LHCb-PAPER,LHCb-CONF,LHCb-DP,LHCb-TDR}

\newpage
\centerline
{\large\bf LHCb collaboration}
\begin
{flushleft}
\small
R.~Aaij$^{32}$,
C.~Abell{\'a}n~Beteta$^{50}$,
T.~Ackernley$^{60}$,
B.~Adeva$^{46}$,
M.~Adinolfi$^{54}$,
H.~Afsharnia$^{9}$,
C.A.~Aidala$^{86}$,
S.~Aiola$^{25}$,
Z.~Ajaltouni$^{9}$,
S.~Akar$^{65}$
% ...

{\footnotesize \it

% ...
$^{9}$Universit{\'e} Clermont Auvergne, CNRS/IN2P3, LPC, Clermont-Ferrand, France\\
% ...
$^{25}$INFN Sezione di Milano, Milano, Italy\\
% ...
$^{32}$Nikhef National Institute for Subatomic Physics, Amsterdam, Netherlands\\
% ...
$^{46}$Instituto Galego de F{\'\i}sica de Altas Enerx{\'\i}as (IGFAE), Universidade de Santiago de Compostela, Santiago de Compostela, Spain\\
% ...
$^{50}$Physik-Institut, Universit{\"a}t Z{\"u}rich, Z{\"u}rich, Switzerland\\
% ...
$^{54}$H.H. Wills Physics Laboratory, University of Bristol, Bristol, United Kingdom\\
% ...
$^{60}$Oliver Lodge Laboratory, University of Liverpool, Liverpool, United Kingdom\\
% ...
$^{65}$University of Cincinnati, Cincinnati, OH, United States\\
% ...
$^{68}$Syracuse University, Syracuse, NY, United States\\
% ...
$^{86}$University of Michigan, Ann Arbor, United States, associated to $^{68}$\\
\bigskip
% ...
\medskip
}
\end{flushleft}
\end{document}